\documentclass[twocolumn, aps,graphicx,prb, showpacs]{revtex4}
\usepackage{graphicx}
\usepackage{amssymb}
\usepackage{amsfonts}
\usepackage{amsmath}
\usepackage{color}
\usepackage{ulem}
\usepackage{float}

\begin{document}

\title{Scalable ion-photon quantum interface based on integrated diffractive mirrors}

\author{M. Ghadimi$^1$, V. Bl\={u}ms$^1$, B. G. Norton$^1$,  P. M. Fisher$^1$,  S. C. Connell$^1$, J. M. Amini$^2$,  C. Volin$^2$,  H. Hayden$^2$,  C. S. Pai$^2$, D. Kielpinski$^1$,  M. Lobino$^{1,3}$,  E.W. Streed$^{1,4,\star}$\\
$^1$\textit{Centre for Quantum Dynamics, Griffith University, Brisbane, QLD 4111 Australia} \\
$^2$\textit{Georgia Tech Research Institute, Atlanta, GA, 30318 USA}\\
$^3$\textit{Queensland Micro and Nanotechnology Centre, Griffith University, Brisbane, QLD 4111 Australia}\\
$^4$\textit{Institute for Glycomics, Griffith University, Gold Coast, QLD 4222 Australia}\\
\textit{$^\star$e.streed@griffith.edu.au}}

\begin{abstract}
Quantum networking links quantum processors through remote entanglement\cite{Kimble2008} for distributed quantum information processing (QIP) and secure long-range communication. Trapped ions are a leading QIP platform, having demonstrated universal small-scale processors \cite{Harty-Lucas-2014} and roadmaps for large-scale implementation \cite{Monroe-Kim-2014}. Overall rates of ion-photon entanglement generation, essential for remote trapped ion entanglement\cite{teleport}, are limited by coupling efficiency into single mode fibres\cite{Hucul-2015} and scaling to many ions. Here we show a microfabricated trap with integrated diffractive mirrors that couples 4.1(6)\% of the fluorescence from a $^{174}$Yb$^+$ ion into a single mode fibre, nearly triple the demonstrated bulk optics efficiency\cite{Hucul-2015}. The integrated optic collects 5.8(8)\% of the $\pi$ transition fluorescence, images the ion with sub-wavelength resolution, and couples 71(5)\% of the collected light into the fibre. Our technology is suitable for entangling multiple ions in parallel and overcomes mode quality limitations of existing integrated optical interconnects \cite{VanDevender-10, Clark-Stick-2014, Merrill-Slusher-2011}. In addition, the efficiencies are sufficient for fault tolerant QIP\cite{Steane-07}. 
\end{abstract}

\maketitle

{\fontsize{9.5pt}{9.5pt}\selectfont

\begin{figure*}
\centering
\includegraphics[width=\textwidth]{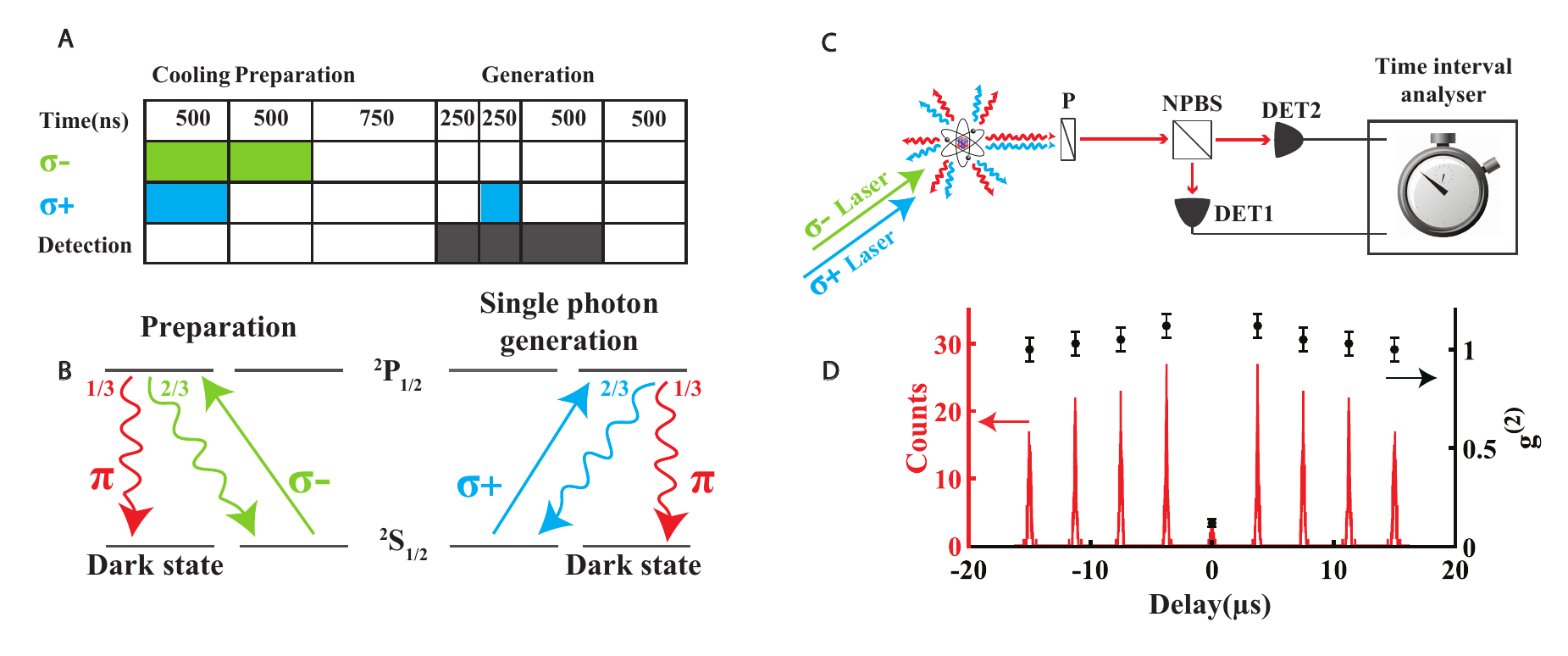}
\caption{Single photon generation protocol. \textbf{A.} Pulse sequence for $\sigma^+$ and $\sigma^-$ lasers and detection gating.  \textbf{B.}  $^{174}$Yb$^+$ level diagram showing $\pi$ photons result in optical pumping to a dark state. \textbf{C.} Experimental set-up of the second order correlation measurement $g^{(2)}$. Ion fluorescence is analysed with a linear polariser (P) to filter out $\sigma$ photons, leaving only $\pi$ photons to be divided by a 50/50 non-polarizing beamsplitter (NPBS) between two photomultiplier tube detectors (DET1, DET2). Arrival time statistics are accumulated by a digital interval analyser. \textbf{D.} Measured coincidence counts and second-order correlation $g^{(2)}$. Peak spacing corresponds to 3.25 $\mu$s experimental repetition period.}
\label{figure2:}
\end{figure*}

A small chain of trapped ions, confined along the node of an oscillating electric field in a Paul trap, provides a well controlled quantum system that can be cooled to the quantum ground state and precisely manipulated with lasers and microwaves. The ions are simultaneously strongly coupled to each other through the Coulomb force, and decoupled from the surrounding environment. The strong mutual coupling is critical for implementing deterministic multi-qubit entangling gates\cite{Cirac-QCIons,MolmerSorensen} while the external decoupling enables memory coherence times approaching one minute \cite{Harty-Lucas-2014} and single qubit gate error rates\cite{Brown-Wineland-2011} below $10^{-4}$. Recent developments include implementation  of the Shor factoring algorithm\cite{Blatt-Shor} and a programmable quantum computer module based on five ions \cite{Debnath-Monroe-2016}. In these experiments the ions' fluorescence was collected using complex multi-element bulk optics, which are unsuitable for scaling to massively parallel systems.

Ion fluorescence plays two complementary roles in quantum information processing: state readout through the collection of multiple photons, and the creation of remote entanglement through entanglement swapping of photons coupled into single optical modes. The fidelity and readout speed for qubit state detection in trapped ion QIP depends only on the fluorescence collection efficiency~\cite{Wolk-2015}. While efficient coupling into single-mode structures, including optical fibres or arrays of waveguides for collection from multiple ions~\cite{Kielpinski2015}, requires a collection apparatus that is also capable of producing a high quality ion image. Collection efficiencies in free space of up to 54.8\% \cite{Maiwald-Leuchs-Largest-Collection-Smallest-Spot-Trapped-Ion-2012} have been realised in custom fabricated parabolic mirror with very large solid angle coverage. However these devices are highly labour intensive to construct and not scalable. A more scalable approach used reflected curved surface optics into microfabricated ion traps\cite{Merrill-Slusher-2011}, but the ion image quality of such system remains insufficient for good single mode coupling. In fact for both local processing and remote networking a scalable optical interface that can efficiently interface multiple ions with single mode guiding structures is necessary to achieve large, massively parallel implementations.

\begin{figure}[H]
\centering
\includegraphics[width=89 mm]{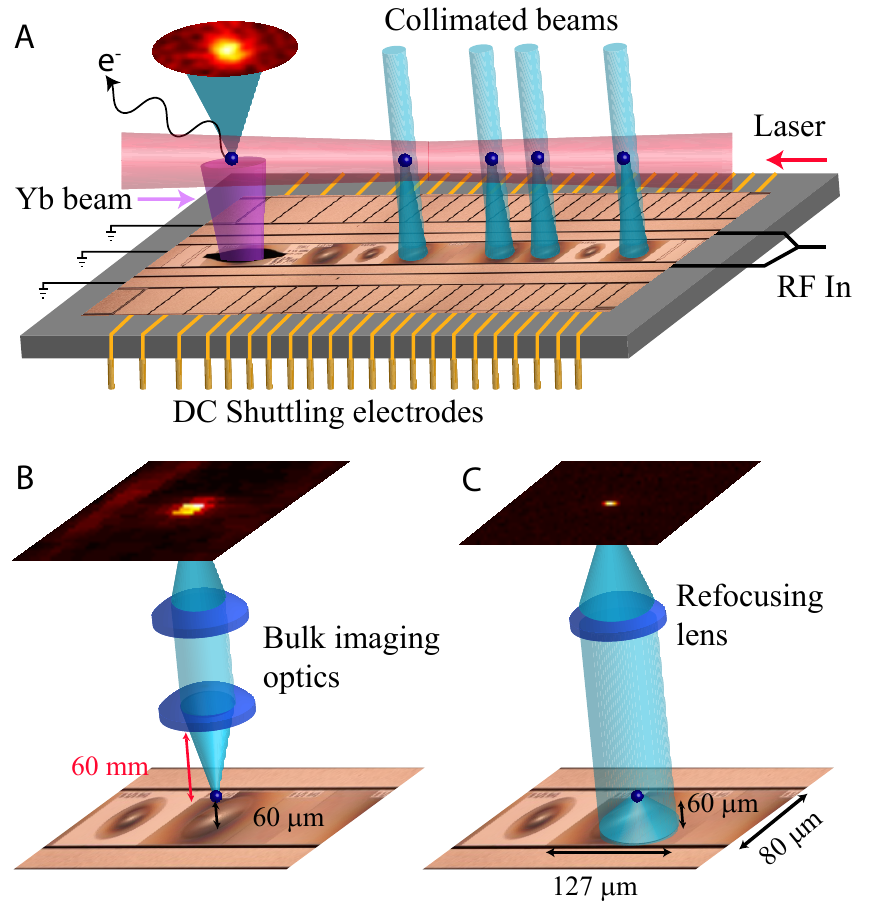}
\caption{\textbf{A.} Schematic of the surface ion trap modified with an array of integrated monolithic diffractive mirrors. Neutral Yb from a thermal beam (purple, left) is photo-ionised in a loading zone, laser cooled, and the $^{174}$Yb$^+$ ion (blue dots) is shuttled to one of the collimator sites by modulating the DC electrode array. Illustrated laser beam (pink) orientation is for clarity, actual is coplanar to chip surface and $45^{\circ}$ to the RF rails. \textbf{B.} Conventional ion fluorescence imaging using external bulk optics with ~1\% collection efficiency. \textbf{C.} Diffractive mirror imaging with 5.8(8)\% collection efficiency of the $\pi$ transition fluorescence.}
\label{figure1:}
\end{figure} 

Diffractive optics offers a solution to this problem since they are scalable, can engineer out geometrical aberrations at the design phase for efficient coupling into single mode structures, and are compatible with several platform such as neutral atom chip traps \cite{atom_chip} or crystal colour centres \cite{NV_center_SIL}. Previous efforts using diffractive lenses with traditional macro traps have demonstrated a 4.2\% collection efficiency~\cite{Streed-11} in free space as well as near diffraction-limited imaging in both fluorescence~\cite{Jechow-11} and absorption~\cite{Streed-Kielpinski-2012-Single-atom-absorption-imaging} modalities, the latter being important for implementing quantum photonic receivers.

Here we exploit the potential of diffractive optics and demonstrate a scalable photon-ion interface realised on a multi-zone micro-fabricated surface trap with integrated diffractive mirrors. Our trap is based on a proven design~\cite{Shappert-Harter-2013} and is shown in Fig~\ref{figure1:}A. The substrate under the central grounding electrode was patterned and etched before coating with 100~nm of aluminium to also act as an array of diffractive mirrors (see Methods). We tested the properties of the integrated mirror by measuring its collection efficiency with a protocol for the creation of triggered single photons and acquiring a near diffraction-limited image of a single ion. With such image quality we were able to obtained an overall coupling efficiency from ion into single mode fibre which doubled what was previously achieved with micro-trap and multimode fibres by direct collection\cite{VanDevender-10} and with the use of diffractive lenses\cite{Clark-Stick-2014} respectively. 

The diffractive mirror has a focal length of 59.6~$\mu$m which correspond to the ion height necessary for high resolution imaging when the collimated flourescence is refocused by an external lens at the desired magnification (Fig~\ref{figure1:}C). The integrated optic has a design diffraction efficiency of 50\%, which includes 92\% aluminum reflectivity in the UV, and is 80~$\mu$m wide by 127~$\mu$m long. The width of the mirror is limited by the centre ground electrode size and the length is set to match the industry standard 127 $\mu$m pitch of V-groove fibre arrays for simultaneous coupling of multiple ions. The diffractive mirror has a numerical aperture of 0.55 in the 80~$\mu$m direction and 0.73 in the 127~$\mu$m direction parallel to the RF rails, capturing 13.3\% of the total solid angle. This is equivalent to a circular NA of 0.68, which maximises the overall rate of entanglement generation per unit surface area of the trap~\cite{Kielpinski2015}.

The trap is loaded with $^{174}$Yb$^+$ ions by isotope selective photo-ionisation from an effusive Yb beam passing through a slot in the chip surface (Fig~\ref{figure1:}A farthest left ion image). Trapped ions are Doppler cooled and shuttled from the loading zone to the various diffractive optic focus sites by varying the potential in the segmented DC electrode arrays. A conventional bulk optics imaging system (Fig~\ref{figure1:}B) allows the ion to be observed at all points along the RF rail for diagnostic purposes. As an indication of the robustness of diffractive optics as an imaging solution, a test diffractive mirror patterned around the Yb oven loading slot was successfully used to image ions, despite the  large central void which reduced its collection efficiency. A magnetic field coplanar with the chip, at $45^{\circ}$ to the RF rails, sets the direction of the quantisation axis. The laser's direction is aligned with the magnetic field such that circularly polarised light excites either the $\sigma^+$ or $\sigma^-$ transitions.

We measured the collection efficiency of the integrated optics using a single photon generation protocol based on optical-pumping (see Fig~\ref{figure2:} and Methods). The protocol relies on selective illumination with $\sigma^+$ and $\sigma^-$ polarised lasers followed by emission of a $\pi$ photon which pumps the atom into a dark ground state (Fig~\ref{figure2:}A,B). Differences in the radiation distribution patterns results in our optic collecting 17.4\% of the fluorescence from $\pi$ polarised and 11.3\% from $\sigma$ polarised transitions. In the direction perpendicular to the quantisation axis and parallel to the collection optic axis, $\pi$ and $\sigma$ photons are orthogonally polarised. With this geometry single $\pi$ photons were selectively detected placing a linear polariser before a photomultiplier tube (PMT). To verify the integrity of our single photon generation protocol we measured the second order correlation $g^{(2)}(\tau)$ of the fluorescence collected through the integrated mirror with a 50/50 non-polarising beam splitter and two detectors (Fig~\ref{figure2:}C). Our large numerical aperture results in polarisation blurring~\cite{Blinov-04} of the $\sigma$ photons far from the collection optic axis. To reduce the transmission of $\sigma$ photons through the polariser we used an iris to temporarily decrease the NA from 0.68 to 0.48 and measured a $g^{(2)}(0)$ of 0.12(2)  (Fig~\ref{figure3:}D). This value is well below threshold of 0.5 for a single photon emitter\cite{quantumth}, but larger than our expected residual value for $g^{(2)}(0)$ of 0.069. The difference is likely due to residual polarisation blurring, imperfections of the polarisation purity of the excitation laser and background scattering of the incident laser light off the chip's surface. The 308~kHz repetition rate of  this protocol is comparable with the present state of the art for remote entanglement ion trapping experiments\cite{Hucul-2015}.

In order to measure the collection efficiency of the diffractive mirror, we ran the triggered single photon protocol 184,000 times and counted 770 photons. Adjusting for known loss processes (i.e. 19\% detector quantum efficiency, 50(5)\% transmission through the iris, and 24\% losses through other optical elements including the polariser) we measured a 5.8(8)\% collection efficiency for our diffractive mirror on the $\pi$ transition. In comparison, our theoretical efficiency of 8.7\% is the product of the 17.4\% solid angle coverage for the $\pi$ transition and the 50\% expected diffraction efficiency. This value indicates that our diffraction efficiency is 33(7)\% and not the expected 50\%, most likely due to fabrication imperfections in the optics.

% dropped center
\begin{figure}[H]
\centering
  \includegraphics[width=80 mm]{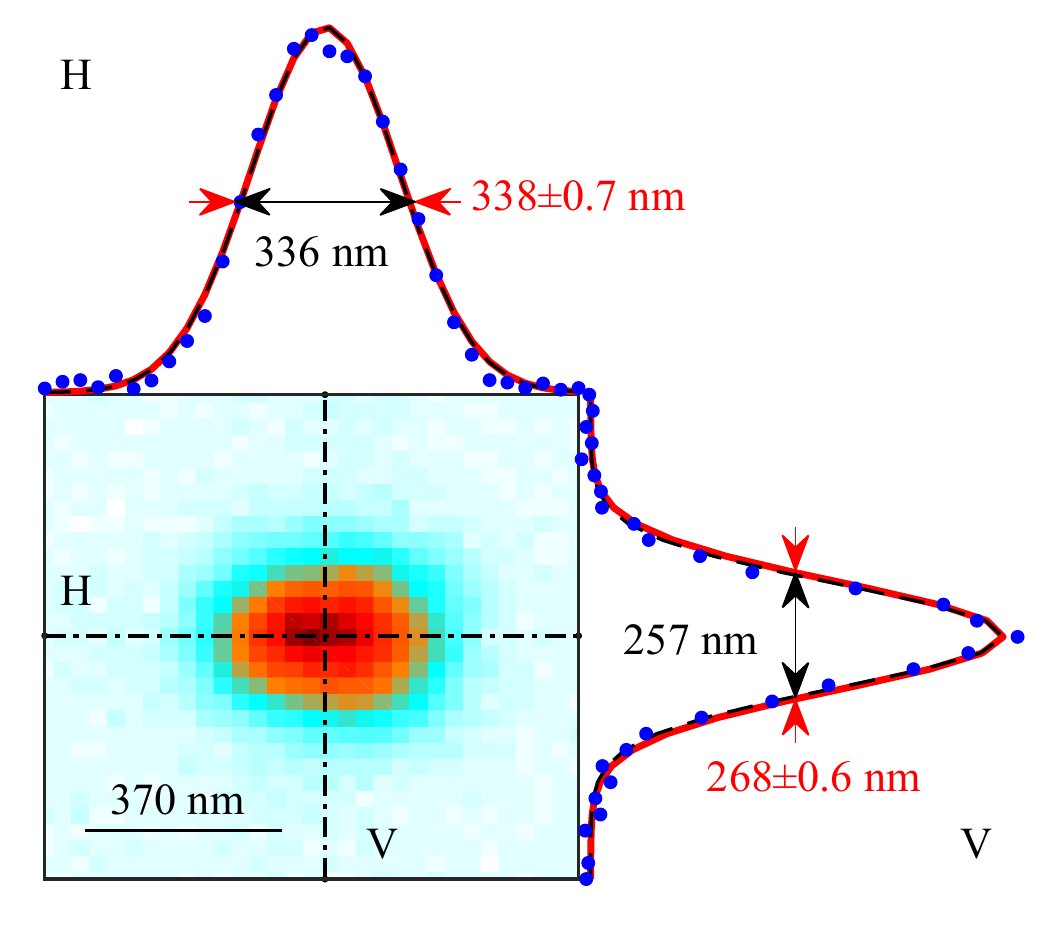}
  \caption{Sub-wavelength image of an  $^{174}$Yb$^+$ ion by an integrated diffractive mirror. Near diffraction limited performance was obtained from a rectangular optic 80$\mu$m  x 127 $\mu$m operating at  NA= 0.56 x 0.73 covering a solid angle $\Omega/4\pi$=13.3\%. Imaged at magnification 390$\times$, 1~s exposure. Horizontal (H) and Vertical (V) cross sections indicate image data (blue points), 338(0.7)~nm x 268(0.6)~nm FWHM diameter gaussian fits (red), and predicted diffraction limit (black dashed).}
  \label{figure3:}
\end{figure}

Figure~\ref{figure3:} shows a near diffraction-limited image of an ion with a sub-wavelength spot size of 338(0.7)~nm horizontal (H) and 268(0.6)~nm vertical (V) FWHM diameter. This is only slightly larger (0.6\% for horizontal and 4\% for vertical) than the diffraction limited size of 336~nm (H) and 257~nm (V), and approaching the state of the art 280 nm circular spot size reported for a speciality 81\% solid angle coverage parabolic reflector optic\cite{Maiwald-Leuchs-Largest-Collection-Smallest-Spot-Trapped-Ion-2012}. The $M^2$ mode quality of the spot was 1.36 (H) and 1.54 (V), compared to an ideal Gaussian ($M^2$=1). This non-ideal behaviour is driven by increased divergence in the beam due to astigmatism from the aperture and the fundamentally non-Gaussian distribution of the ion's emission.

We benchmarked the collection and imaging capabilities of the integrated mirror by coupling the ion's fluorescence into a single mode fibre. We remove the iris and polariser from the set-up and used a mode matching telescope for adjusting the average radius of the ion fluorescence image to that of a single fibre mode with an estimated spatial mode overlap of 98\%. The ion image's average $M^2$ of 1.45 reduces the predicted coupling efficiency to 68\%. We measured a transmission from the ion through a single mode fibre of 57(4)\% which combined with a 80\% fibre transmission due to propagation and Fresnel losses correspond to a single mode coupling efficiency of 71(5)\%. This value is in good agreement with the estimated 68\% coupling efficiency and combined with the 5.8(8)\% $\pi$ transition collection efficiency from the diffractive optics and 8.3\% losses through other optical elements, gives us a total measured coupling efficiency from ion to fibre of 4.1(6)\%. This ion-fibre coupling efficiency is nearly triple the previous best of 1.4\% using a conventional lens~\cite{Hucul-2015}, corresponding to a 8.6 times gain in entanglement generation rate. This efficiency could be substantially improved with larger NA optics or modifying the optic's design to mode-match a specific transition's intensity distribution to a single mode fibre. More sophisticated multi-level diffraction gratings\cite{Grating-99} could improve the diffraction efficiency towards 99\%.

We have realised a scalable architecture for interfacing a single ion with single mode optical fibre and free space based on integrated diffractive mirrors.  Using a triggered single photon generation protocol we measured a collection efficiency for the integrated optics of 5.8(8)\% a coupling efficiency into a single mode optical fibre of 71(5)\%. These mirrors are monolithic with the metallic trap electrode, cover a large solid angle, and can be within a few tens of microns from the ion without distorting the RF trapping potential. They are therefore an ideal platform for the implementation of quantum networks and remote entanglement sharing with trapped ions but also with other quantum light sources such as neutral atom chip traps \cite{atom_chip} or fixed emitters such as crystal colour centres like NV$^-$in diamond \cite{NV_center_SIL}.

\section*{Methods}
\fontsize{8pt}{8pt}\selectfont
{\bf Trap Fabrication}

The micro-fabrication procedure\cite{Shappert-Harter-2013} for the surface trap includes depositing multiple layers of aluminium (R=92\%, $\lambda$=370~nm) separated by silicon dioxide insulating layers of various thicknesses. In order to incorporate the diffractive optical elements into this procedure, the oxide surface separating the control electrodes from the ground layer was patterned using e-beam lithography and reactive ion etching of the oxide. The patterned area was subsequently metallised with a 100~nm thick layer of aluminium and the chip was inspected to ensure that the contours of the diffractive optic had not compromised the electrical integrity of the ground electrode. The groove step design is a hybrid of a four level grating near the centre with low spatial period and a two level grating near the edges where there is a high spatial period. Grating height step size is 45~nm for 4 level area and 90~nm for 2 level area. Approximating small chunks of the grating as 1D structures, the blazing profile was optimised in GSolver to account for the finite height of the grating and vector diffraction effects.

Fabrication imperfections can cause a mismatch between the optic focal point and the expected RF node altitude of $58.6~\mu$m above the surface of the trap. This is a critical matching problem due to the small depth of focus in low aberration, large aperture ion imaging \cite{Jechow-11}. To guard against this possibility our five collimating optics were fabricated with focal lengths ranging from $f=58.6~\mu$m to $62.6~\mu$m in 1$\mu m$ steps. Stray electrical fields from the neighbouring oven loading zone precluded the use of the nominal $f_0$=58.6~$\mu$m mirror site and instead experiments were performed on the $f_{+1 \mu m}$=59.6~$\mu$m focal length collimator. A DC potential was applied to shift the ion off the RF node and position it at the focal point. We observed it was more reliable to pull the ion towards the surface of the trap using a DC potential, rather than pushing it away the trap. While we did not observe a mismatch, for higher numerical apertures this parameter will become even more stringent. In future iterations this could be corrected dynamically with additional RF lines or statically by laser trimming the RF electrodes.

{\bf Triggered Single Photon Generation Protocol}

To measure the collection efficiency of the integrated mirror we implemented an optical pumping based single photon generation protocol (Fig \ref{figure2:}). The protocol relies on decay into a dark optically pumped state requiring the emission of a terminal $\pi$ polarised photon rather than a $\sigma$ polarised photon (Fig \ref{figure2:}B). The ion was first Doppler cooled for 500 ns with two laser beams of $\sigma^+$ and $\sigma^-$ polarisations with 7~$\mu$W and 100~$\mu$m diameter at 370~nm, detuned -10~MHz from resonance. Each scattered photon has a 2/3 chance of returning to its original ground state via a $\sigma$ decay, or a 1/3 chance of being optically pumped into the other ground state via a $\pi$ transition. 
The ion was then optically pumped in the $^2 S_{1/2}$ $m_F=+1/2$ state by 500~ns illumination with just the $\sigma^-$ light. 750 ns of wait time was introduced for acousto-optic modulator switching time to ensure all laser beams were off. The detection window was then activated for 1000 ns with a 250 ns pulse of $\sigma^+$ occurring 250 ns into the detection window.  On average this process takes three scattering events (2 $\sigma$ and 1 $\pi$) and the whole duration of one cycle is 3.25~$\mu s$. Since the collection optic was oriented perpendicular to the quantisation axis, the $\sigma$ and $\pi$ photons had perpendicular linear polarisations allowing us to filter out the single $\pi$ photon with a linear polariser and detect it on the PMT. Occasional decays (0.5\%) into the $^2D_{[3/2]}$ dark state were repumped out with 650~$\mu$W of 935~nm light and did not have a meaningful impact on the experiment.

\section*{Author contribution statement}

Ion trap construction and experimental work was performed by M. Ghadimi, V. Bl\={u}ms, B. G. Norton, P. M. Fisher, S. C. Connell, and D. Kielpinski with supervision by D. Kielpinski, M. Lobino, and E.W. Streed. The chip trap was fabricated by J. M. Amini, C. Volin, H. Hayden, and C. S. Pai at Georgia Tech Research Institute. The manuscript was prepared by M. Ghadimi, M. Lobino, and E.W. Streed in consultation with the other authors.

\section*{Acknowledgements}
This material is based upon work supported by the Office of the Director of National Intelligence (ODNI), Intelligence Advanced Research Projects Activity (IARPA) under U.S. Army Research Office (ARO) contract number W911NF1210600. All statements of fact, opinion, or conclusions contained herein are those of the authors and should not be construed as representing the official views or policies of IARPA, the ODNI, or the U.S. Government. This work was supported by the Australian Research Council (ARC) under
DP130101613. D.K. was supported by ARC Future Fellowship FT110100513. E.W.S was supported by ARC Future Fellowship FT130100472. M.L. was supported by ARC-Decra DE130100304. The phase Fresnel mirrors were fabricated by M. Ferstl at the Heinrich-Hertz-Institut of the Fraunhofer-Institut f\"{u}r Nachrichtentechnik in Germany.

\section*{Competing Interests} The authors declare that they have no competing financial interests.

\section*{Correspondence} Correspondence and requests for materials should be addressed to EWS ~(email: e.streed@griffith.edu.au).

\end{document}